\documentclass[prd,notitlepage,superscriptaddress]{revtex4-1}
\usepackage{latexsym,amssymb,amsfonts,amsmath}

\usepackage[dvips]{graphicx}

\newcommand{\la}{\langle}
\newcommand{\ra}{\rangle}
\newcommand{\nnn}{{\cal N}}

\newcommand{\eq}{\begin{equation}} 
\newcommand{\eqx}{\end{equation}}
\newcommand{\en}{\begin{enumerate}} 
\newcommand{\enx}{\end{enumerate}}

\newcommand{\bi}{\begin{itemize}} 
\newcommand{\ei}{\end{itemize}} 

\newcommand{\s}{\varsigma}

\newcommand{\f}[2]{\frac{#1}{#2}}
\newcommand{\lra}{\longrightarrow}

\newcommand{\ka}{\kappa}
\newcommand{\lam}{\lambda}
\newcommand{\xpr}{\vec b}

\newcommand{\alp}{\alpha'}

\newcommand{\bt}{\beta}
\renewcommand{\th}{\theta}

\newcommand{\dl}{\delta}

\newcommand{\rr}[4]{#1, {\it #2 \/}{\bf #3} #4}
\newcommand{\rrs}[3]{{\it #1 \/}{\bf #2} #3}
\newcommand{\rra}[2]{{#1}, {#2}}

\renewcommand{\Im}{{\rm Im}}

\newcommand{\W}{{\cal W}}

\begin{document}

\title{High-energy bounds on total cross sections in  $\nnn\!=\!4$
	SYM from AdS/CFT}

\author{M.~Giordano}
\email[E-mail address: ]{matteo.giordano@cea.fr}
\affiliation{Dipartimento di Fisica, Universit\`a di Pisa,\\
Largo Pontecorvo 3, I-56127, Pisa, Italy}
\affiliation{Institut de Physique Th\'eorique  CEA-Saclay \\ F-91191
Gif-sur-Yvette Cedex, France}

\begin{abstract}

Using the AdS/CFT correspondence, we study the high-energy behavior
of scattering amplitudes in {$\nnn\!=\!4$} SYM gauge theory
for dipole-dipole {\it soft} elastic scattering, described
in the Wilson-loop correlator formalism. The amplitudes are evaluated
in the dual picture, at large impact-parameter, by considering the
exchange of supergravity fields between certain minimal surfaces in
Euclidean $AdS_5$, and performing the appropriate analytic
continuation to Minkowskian signature. The purely elastic behavior at
large impact-parameter is then combined with the unitarity constraint
in the central region, in order to derive an absolute bound on the
high-energy behavior of the dipole-dipole total cross section. The
possibility to obtain a stronger bound by assuming a larger domain of
validity for the AdS/CFT result is also discussed.
\end{abstract}

\maketitle

\section{Introduction}

High-energy {\it soft} hadron-hadron scattering is
known to be one of the hardest open problems of strong
interactions. Indeed, since the momentum transfer is small (typically
$\sqrt{|t|}\lesssim 1 {\rm GeV}$), 
this kind of processes involves the
nonperturbative,  strong coupling regime of the underlying microscopic
theory, namely Quantum Chromodynamics. Nevertheless, 
a few results can be obtained using the fundamental properties of a
consistent quantum field theory, established long ago in the $S$-matrix
formalism: {\it unitarity}, coming from the
conservation of probabilities, and {\it analyticity}. These
properties, combined with  the existence of a ``mass gap'' in the
asymptotic particle spectrum, 
lead to the celebrated Froissart bound for the total cross section
\cite{Froissart},  
which corresponds (up to logarithms) to an intercept not greater than $1$ 
for the leading Regge singularity, usually called the ``Pomeron''.

In recent years, a new tool to deal with strong coupling physics has
appeared, namely the Gauge/Gravity duality, which has raised the hope
to understand high-energy soft scattering amplitudes in gauge theories
by mapping them into appropriate quantities in the dual gravity theory.
The first realization of the duality is the well known AdS/CFT
correspondence~\cite{adscft}, which relates
type IIB string theory on $AdS_5\times S^5$ in the weak-coupling,
supergravity limit to $\nnn=4$ SYM theory at strong coupling and large
number of colours. Since this theory is a conformal field theory and thus
non-confining, its behaviour is expected to differ from that of a
confining theory like QCD. However, attacking the problem of soft
amplitudes in this context may be a useful laboratory 
for further developments in QCD. Indeed, the study of  soft high-energy
scattering amplitudes in  $\nnn=4$ SYM using the AdS/CFT
correspondence, and more generally Gauge/Gravity duality, has
attracted much attention in the 
literature~\cite{Jani,Jani1,Brow0,Corn1,Tali1,Muel,LP,Khar}. 

In the conformal case there is no mass gap, and thus the Froissart bound is
not expected to be valid for $\nnn =4$ SYM theory. However, unitarity and
analyticity are still expected to 
hold, and so it is interesting to examine the question of high-energy bounds
in this context. In~\cite{adsbounds} we have used unitarity,
analyticity and the AdS/CFT correspondence to give a precise account
of soft high-energy elastic amplitudes in the $\nnn=4$ supersymmetric
gauge theory. In particular, we have combined the
knowledge obtained from AdS/CFT in the region of applicability
of the supergravity approximation, $i.e.$, the large impact-parameter
region where the amplitude is essentially elastic, with the
constraints coming from analyticity and  unitarity, in order to obtain
high-energy bounds on total cross sections. More precisely, the
following ingredients have been used.
\begin{enumerate}
\item The r{\^o}le of massive quarks and antiquarks ($Q,\bar Q$)
  in the AdS/CFT correspondence is played, as in \cite{Wilson}, by
  the massive $W$ bosons arising from breaking $U(N+1)\rightarrow
  U(N)\times U(1),$ where one brane is considered away from the $N\to
  \infty$ others. In turn, the r\^ole of  hadrons is devoted to
  ``onia'' defined as linear combinations of $Q \bar Q$ colorless
  ``dipole'' states \cite{Muel1} of average transverse
  size $\la{|\vec{R}|}\ra$, which sets the scale for the onium mass.

\item The dipole amplitudes are obtained using the Wilson loop
  formalism~\cite{Nacht,Nachtr} in Euclidean space, in order to avoid the
  complications of the Lorentzian AdS/CFT correspondence (see
  $e.g.$~\cite{lorads}). Performing the appropriate analytic continuation
  \cite{Megg,crossing}, one obtains the
  physical amplitude in Minkowski space. 

\item The AdS/CFT correspondence relates the calculation of
  the Euclidean Wilson loop correlator to a {\it minimal surface} problem in
  the AdS bulk, which has been solved \cite{Jani} at large
  impact-parameter distance $L$ (beyond the Gross-Ooguri transition
  point~\cite{Gross}) using the knowledge of
  (quasi-)disconnected minimal surfaces, connected by the exchange of
  supergravity fields propagating in the bulk.
\end{enumerate}
After analytic continuation, one obtains the scattering
amplitude in the impact-parameter representation at large impact
parameter; combining this with unitarity to fix a bound in
the lower impact-parameter domain, it leads to a determination of
new energy bounds on the forward elastic amplitudes, or, equivalently, 
on the total cross sections in $\nnn=4$ SYM.

\section{Elastic amplitudes from Euclidean Wilson Loop correlators}

\begin{figure}[t]
  \centering
  \includegraphics[width=0.41\textwidth]{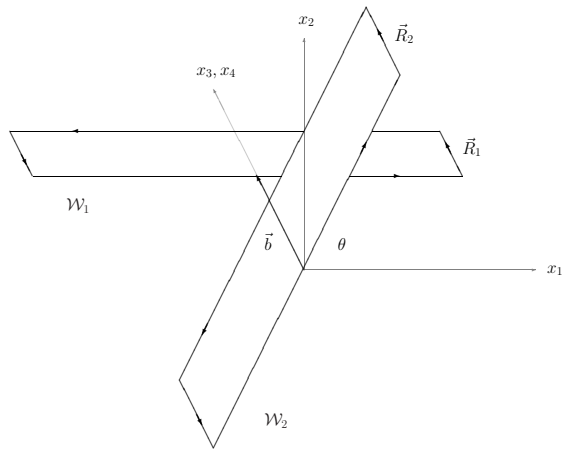}
  \caption{{\it Geometry of the Wilson loops in Euclidean space.} 
    The transverse kinematic variables ($\vec b,\vec R_{1,2})$ remain
    unchanged by the analytic continuation to Minkowski space, while $\th \to
    -i\chi$, see text.} 
  \label{1}
\end{figure}

In the eikonal approximation, dipole-dipole elastic scattering
amplitudes in the {\it soft} high-energy regime can be conveniently
expressed in terms of the normalized connected correlator ${\cal
  C}_M$ of Wilson {\em loops} in Minkowski space~\cite{Nachtr}
\eq
\label{e.ampinit}
{\cal A }(s,t;\vec{R}_1,\vec{R}_2)= -2is \int
d^2\xpr\, e^{i{\vec q}\cdot\xpr}\, {\cal C}_M(\chi,\xpr;\vec{R}_1,\vec{R}_2)
\equiv -2is \int d^2\xpr\,  e^{i{\vec q}\cdot\xpr}
\left[\f{\la\W_1\W_2\ra}{\la\W_1\ra\la\W_2\ra}-1\right], 
\eqx
where $t=-\vec{q}^{\,2}$, $\vec{q}$ being the transverse transferred
momentum (here and in the following we denote with
$\vec{v}$ a two-dimensional vector), and the Wilson loops follow the
classical straight-line trajectories for quarks (antiquarks, in
parenthesis) \cite{Verl}: 
\eq
\W_1\lra X_1^\mu=b^\mu+ u_1^\mu\tau\,(+r_1^\mu)\ ; \quad \quad \W_2\lra
X_2^\mu=u_2^\mu\tau\, (+r_2^\mu)\ .
\label{traj}
\eqx
Here $u^\mu_{1,2}$ are unit
time-like vectors along the directions of the momenta defining the
dipole classical  
trajectories, and moreover $b^\mu=(0,0,\vec{b})$ and $r^\mu_{1,2} =
(0,0,\vec{R}_{1,2})$, with $R_i=|\vec{R}_i|$ the quark-anti\-quark transverse
separation. The loop contours are then closed at positive
and negative infinite 
proper-time $\tau$ in order to ensure gauge-invariance. 
The scattering amplitude for two onium states can then be
reconstructed from the dipole-dipole amplitude after folding with the
appropriate wave-functions for the onia. 
The mass of an onium state 
is expected to be of the order of the inverse of its
average radius, $m \sim\la R \ra^{-1}$.
The geometrical parameters of the configuration can be related to the
energy scales by the relation $\cosh \chi\equiv s/2m^2-1\equiv \s -1$, which at
high energy reads $\chi\mathop\sim  \log\s$, where $\chi$ 
is the hyperbolic angle 
(rapidity) between the trajectories of the dipoles, $Y=\log\s$ the total rapidity
and $m$ the masses of the two onia, taken to be equal for simplicity.

As already mentioned in the Introduction, it is convenient to
exploit the Euclidean version of the correspondence, and then to 
reconstruct the relevant correlation function ${\cal C}_M$ from its
Euclidean counterpart ${\cal C}_E$ by means of analytic
continuation~\cite{Megg,crossing}. The Euclidean
approach has already been 
employed in the study of high-energy soft scattering amplitudes by
means of non perturbative
techniques~\cite{Jani,Jani1,ILM}, including numerical 
lattice calculations~\cite{Latt}.
The Euclidean normalized connected correlation
function is defined as
\eq
  \label{eq:sugraexch0}
  {\cal C}_E(\theta,\xpr;\vec{R}_1,\vec{R}_2) \equiv  \frac{\la \W_1
    \W_2 \ra}{\la \W_1 \ra \la \W_2 \ra} - 1\ , 
\eqx
where $\W_i$ are now Euclidean Wilson loops evaluated along the
straight-line paths $\W_1\lra X_1^\mu=b^\mu+u_1^\mu\tau\,(+r_1^\mu)$ and $\W_2\lra
X_2^\mu=u_2^\mu\tau\, (+r_2^\mu)$, closed at infinite proper
time, see Fig.~\ref{1}. The variables $b$ and $r_i$ are the same
defined above in the Minkowskian 
case (we take Euclidean time to be the first coordinate to
  keep the notation close to the Minkowskian case).
Here $u_1$ and $u_2$ are unit vectors forming an angle $\theta$
in Euclidean space.

The physical correlation function ${\cal C}_M$ in Minkowski space is
obtained by means of the analytic continuation~\cite{Megg,crossing}
\eq
\label{e.ancont}
{\cal C}_M(\chi,\vec{b};\vec{R}_1,\vec{R}_2) =
{\cal C}_E(-i\chi,\vec{b};\vec{R}_1,\vec{R}_2),\quad \chi\in\mathbb{R}^+ \ ,
\eqx
where the analytic continuation of ${\cal C}_E$ is performed starting from 
the interval $\theta\in (0,\pi)$ for the Euclidean angle (the
restriction on the range of $\chi$ and $\theta$ does not imply any
loss of information, due to the symmetries of the two theories).

\section{Wilson loop correlators from AdS/CFT}

Within the AdS/CFT correspondence, the correlators of Wilson loops
in the gauge theory, such as those of 
Eq.~\eqref{e.ampinit}, are related to a minimal surface in the bulk of
$AdS_5$ having as boundaries the two Wilson loops, corresponding to
the minimization of the Nambu-Goto action. 
When $L\equiv \vert \vec b \vert \lesssim R_1,R_2 $  there exists a connected minimal
surface with the sum of the two loops as its disjoint boundary (see
$e.g.$~\cite{Zarembo}), although its explicit expression is 
difficult to obtain. However, when $L\gg R_1,R_2$ the solution
simplifies, and the minimal surface
has two independent, (quasi-)disconnected components: in order to calculate
the correlator one then exploits, as in Ref.~\cite{MaldCor}, the
explicit solutions corresponding to the two loops connected by the
classical supergravity interaction, $i.e.$, by the exchange between them (see
Fig.~\ref{2}-1) of the lightest fields of the $AdS_5$
supergravity, namely the graviton ($G$), the anti-symmetric tensor ($B$), the
dilaton ($D$), and the (tachyonic) Kaluza-Klein (KK)
scalar mode ($S$). This is the case we have considered in~\cite{adsbounds}, using the large-$L$ 
behavior of the dipole-dipole impact-parameter amplitude evaluated
in~\cite{Jani}, with a slight generalization to unequal dipole sizes.

For $L\gg R_1,R_2$, and in the weak gravitational field domain, the
Euclidean normalized connected correlation function has the form  
\eq
  \label{eq:sugraexch}
  {\cal C}_E  
  =  \exp{\left(\sum_\psi \tilde{\dl}_\psi\right)} -1\ , \qquad
  \tilde{\dl}_\psi \equiv  \frac{1}{4\pi^2 \alpha'{}^2} 4 \int d\tau_1 d\tau_2 \,
  \frac{dz_1}{z_{1x}}\frac{dz_2}{z_{2x}} \frac{\delta S_{NG}}{\delta \psi}(\tau_1,z_1)
\  G_\psi(X_1,X_2) \frac{\delta S_{NG}}{\delta \psi}(\tau_2,z_2)\ ,
\eqx
where $S_{NG}$ is the Nambu-Goto action, $\alpha'=1/\sqrt{4\pi
  g_sN_c}$ and  $g_s$ is the string coupling, related to the gauge
theory coupling by $g_{\rm YM}^2 =2\pi g_s$.
Here $\frac{\delta S_{NG}}{\delta \psi}$ is the coupling of the
world-sheet minimal surfaces attached to the two Wilson loops to the supergravity field
$\psi$. Moreover, $\tau_i$ is the proper time on 
world-sheet $i=(1,2)$, and $z_i$ are the fifth coordinates of
points $X_i$ in $AdS_5$, namely 
\eq
  \label{eq:coord}
\begin{aligned}
  X_1&=\left(u_1^\mu\tau_1 + x_1^\mu + b^\mu,z_1\right)\, , 
  && X_2=\left(u_2^\mu\tau_2 + x_2^\mu,z_2\right)\, ,\\
 u_1^\mu &= (1,0,\vec{0})\, ,\quad  u_2^\mu =
  (\cos\theta,-\sin\theta,\vec{0})\, , &&
x_{i}^\mu = \sigma_{i}(z_i) \f{r_i^\mu}{R_i}\, ,\quad
\sigma_i\in[0,R_i]\, ,\quad i=1,2\ ,
\end{aligned}
\eqx
where $\sigma_{i}(z_i)$ is determined by inverting the solution of the
minimal surface equation $z_i=z_i(\sigma_i)$. The derivatives $z_{ix}\equiv
\f{\partial z_{i}}{\partial \sigma_{i}}$ are given by~\cite{Wilson} 
\eq
z_{ix}=\left(\frac{z_{i{\,\rm
      max}}}{z_i}\right)^2\sqrt{1-\left(\frac{z_i}{z_{i {\,\rm max}}}\right)^4}, 
\quad z_{i\,\rm{max}}={R}_{i}\,\frac{[\Gamma(1/4)]^2}{(2\pi)^{3/2}}\ .
\label{minimal}
\eqx 
In Eq.~\eqref{eq:sugraexch}, $G_\psi(X_1,X_2)$ is the Green function
relevant to the exchange of field 
$\psi$, which depends only on invariant bitensors and scalar
functions~\cite{bitensor} of the AdS invariant
\begin{equation}
  \label{eq:AdSinv}
  u = \frac{(z_1-z_2)^2 + \sum_{j=1}^4 (X_1^j-X_2^j)^2}{2z_1z_2}\ .
\end{equation}
Working out the Green functions and the couplings 
corresponding to the exchange of the various supergravity fields, 
it is found that the leading
dependence (the ``leading'' term in $\theta$ is understood as
  the leading term in $\chi$ after analytic continuation, see below)  on
$\theta$, $L$ and  ${R}_{i}$ for the various 
terms of \eqref{eq:sugraexch} is the following: 
\eq
\begin{alignedat}{7}
\label{eq:phaseshifts}
 \tilde{\delta}_S  &=  \kappa_S\ \frac{1}{\sin\theta} 
  \left(\frac{{R}_{1}{R}_{2}}{L^2}\right)&&  \equiv    a_S 
  \ \frac{1}{\sin\theta}\, ; 
 &&\ \ \ \ \tilde{\delta}_D  &&=    \kappa_D \ \frac{1}{\sin\theta}\
  \left(\frac{{R}_{1}{R}_{2}}{L^2}\right)^3&&  \equiv    a_D\
  \frac{1}{\sin\theta}\, ; \\
 \tilde{\delta}_B  &=    \kappa_B\  
\frac{\cos\theta}{\sin\theta}\left(\frac{{R}_{1}{R}_{2}}
{L^2}\right)^2&&  \equiv    a_B
  \ \frac{\cos\theta}{\sin\theta}\, ;
&&\ \ \ \ \tilde{\delta}_G  &&=    \kappa_G\
\frac{(\cos\theta)^2}{\sin\theta}\left(\frac{{R}_{1}{R}_{2}}
  {L^2}\right)^3&&  \equiv    a_G
  \ \frac{(\cos\theta)^2}{\sin\theta}\, , 
\end{alignedat}
\eqx
factorizing
explicitly the angular dependence from the rest. The results of \eqref{eq:phaseshifts} 
correspond to the case in which $\vec{b}$, $\vec{R}_1$ and $\vec{R}_2$
lie along the same direction in the transverse plane. 
The factors $\kappa_{\psi}$ for each supergravity field are numerical
factors of order ${\cal O}(\lambda/N_c^2)$, with $\lambda=g_{\rm
  YM}^2N_c/4\pi$ the 't Hooft coupling, as one expects from the
topology of the configuration. 
Performing now the analytic continuation $\theta\to -i\chi$, leading to the 
phase shifts $i\delta_\psi\equiv{\tilde\delta_\psi}(\theta\to-i\chi)$, one
finally obtains for the Minkowskian correlation function
\eq
\begin{aligned}
  {\cal C}_M = \exp\left(i\sum_\psi \delta_\psi\right)-1\ ,\qquad
  \delta_S = a_S\,\f{1}{\sinh\chi},\quad   \delta_D = a_D\,\f{1}{\sinh\chi},\quad
  \delta_B = a_B\,\f{\cosh\chi}{\sinh\chi},\quad   \delta_G = a_G\,\f{(\cosh\chi)^2}{\sinh\chi}\ .
\end{aligned}
\label{CM}
\eqx
We notice that under crossing, $i.e.$, under $\chi\to i\pi-\chi$~\cite{crossing}, the
phases $\delta_S$, $\delta_D$ and $\delta_G$ are symmetric, while
$\delta_B$ is antisymmetric.

\section{Eikonal amplitude in impact-parameter space}

\subsection{The weak field constraint}
The range of validity of the results above is determined by
requiring~\cite{Jani} that the effect of the gravitational perturbation $\dl
G_{tt}$ generated by each of the string world-sheets on the other one
is smaller than the background metric $G_{tt}$, therefore ensuring that
one is actually 
working in the weak-field limit. 
Considering the effect of world-sheet 2 on world-sheet 1, 
the strongest constraint is obtained from the evaluation of the
maximal gravitational 
field produced at the point $\tau_1=0,$ where the distance between the
loops is minimal. The weak gravitational field 
requirement reads
\eq
\label{e.gttc}
 \f{\dl G_{tt}}{G_{tt}}\ll 1, \quad G_{tt}\equiv\f{1}{z_1^2}\ ,
\eqx
where $G_{tt}$ is the background metric term 
in the Fefferman-Graham parameterization of $AdS_5$. In order to find
the explicit expression  
of condition \eqref{e.gttc}, we note that $\tilde{\dl}_\psi$ in
\eqref{eq:sugraexch} can be also interpreted as 
an integral over the string world-sheet 1  of the
corresponding supergravity field $\psi(X_2)$ 
produced by the other, tilted world-sheet 2, namely 
\eq
\label{e.field}
\delta \psi (X_1(\tau_1,z_1))= \f{1}{2\pi \alp}  \int 2 d\tau_2 \f{dz_2}{z_{2_x}} 
\ G_\psi(X_1,X_2)\ \frac{\delta S_{NG}}{\delta \psi}(\tau_2,z_2)
\ . 
\eqx
Applying this generic equation to the dominant graviton contribution, 
one finds the constraint
\eq
\f{\dl G_{tt}}{G_{tt}} \propto \f{z_1^4 R_2^3}{L^7}(\cos\th)^2 \ll 1\ .
\eqx
This constraint is most restrictive when evaluated at 
$z_1= z_{1\,{\rm max}}\propto R_1$, which
is as far as the string world-sheet extends into the $5${th} dimension
of $AdS_5$. 
Performing the analytic continuation $\theta\to
-i\chi$, and interchanging the r\^oles of the two
world-sheets by switching the subscripts 1 and 2 in the
result above in order to get the maximal constraint, one finally obtains ($\s=s/2m^2-1$)
\eq
\label{e.constraint}
     {L^2} \gg L_{max}^2 \equiv \f {{{R}_{1}{R}_{2}}\s^{\f47}}{\left[{\rm
     min}\left(\sqrt{\frac{{R}_{1}}{{R}_{2}}},\sqrt{\frac{{R}_{2}}{{R}_{1}}}\,\right)\right]^{\f 27}}\ .
\eqx

\subsection{The elastic eikonal hypothesis}

From expression \eqref{e.ampinit} one can determine the
impact-parameter partial amplitude $a(\chi,\xpr)$ corresponding to the
dipole-dipole elastic  amplitude ${\cal A}$, $i.e.$, suppressing the
dependence on the sizes of the dipoles, $a(\chi,\xpr) = -i{\cal
  C}_M(\chi,\xpr)$. 
In the large-$L$ region the AdS/CFT result gives
\eq
\label{e.impact}
a_{tail}(\chi,\xpr) =  i\Big[1-\exp\Big(i\sum_\psi
\dl_\psi\Big)\Big]\ ,
\eqx
with the phase shifts specified by \eqref{CM}.
This expression can be trusted as long as the solution for the
minimal surface problem is disconnected and, as discussed above, as
long as the weak gravitational field 
constraint \eqref{e.constraint} is satisfied. Note that expression
\eqref{e.impact} corresponds to a purely elastic amplitude, in
agreement with the planar limit implied 
by the AdS/CFT correspondence.

Although the result  \eqref{e.constraint} 
expresses a stringent constraint on the 
impact-parameter range due to the weak gravitational field condition required
in applying the AdS/CFT correspondence, 
one can try to extend the results by adopting an $S$-matrix
point-of-view. Indeed, the exponential  form of \eqref{e.impact} is
typical of  a resummation of non-interacting 
($i.e.$, independent) colorless exchanges (on the gauge theory side)
which can be taken into  
account in order to possibly enlarge its domain of validity, assuming
then the validity of the eikonal approximation for a purely
elastic scattering amplitude.  
From the AdS/CFT correspondence point-of-view, we expect 
that the gravitational field is strong for an impact-parameter
distance below the weak-field limit, so inducing
graviton self-interactions which would spoil this independent resummation; 
nevertheless, for completion, we will suppose that the eikonal formalism
for the elastic amplitude may be extended in some larger phase-space
region. We will then examine, 
from the empirical $S$-matrix  point-of-view, 
whether and  down to which value of the impact-parameter separation
the formula \eqref{e.impact} 
could be used beyond the constraint  \eqref{e.constraint}, i.e., 
up to what impact-parameter distance it is mainly
the exchange of  independent gravitons 
which builds the whole amplitude. 

As a first step beyond our AdS/CFT correspondence result, one could infer from an $S$-matrix 
model formulation that the amplitude 
\eqref{e.impact}
is reliable as long as
the dominant graviton-induced phase shift $\dl_G$ is small. Following formulas 
(\ref{CM}), this means that (at large energy)
$\frac{L^2}{{R}_{1}{R}_{2}}
  \gg \left(\kappa_G\s\right)^{\f13}$; 
 more precisely,
 \eq
{L^2}> L^2_{min}\equiv {{R}_{1}{R}_{2}}
\left(\f {\kappa_G}\pi\ \s\right)^{\f 13}
\label{limit}
\eqx
in order for the eikonal formula \eqref{e.impact} to be physically sensible,
 requiring the phase shift
 $\dl_G\le \pi$. This extreme minimal bound ensures that
$\Im\,a_{tail}(\chi,\xpr)$ be not oscillating with $L$, so yielding 
a non-oscillating behavior for the $L$-dependent partial cross
section. Indeed, it is reasonable to expect that more 
and more inelastic channels would open up when going from the peripheral
to the central impact-parameter domain.  
We see that the request of a weak 
gravitational field gives a stronger constraint than the one coming from
the $S$-matrix model point-of-view.

\subsection{Characteristic impact-parameter scales}

\begin{figure}[t]
  \centering
\includegraphics[width=0.55\textwidth]{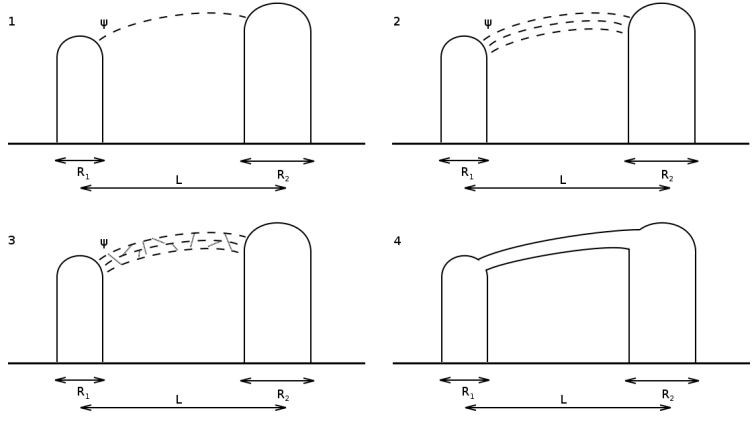}
\caption{{\it Interperetation of the different regimes of supergravity interaction in
    impact-parameter space.} See text.}
\label{2}
\end{figure}

Let us consider  a range of validity
of  \eqref{e.impact} varying from its AdS/CFT determination
\eqref{e.constraint} to its  maximal $S$-matrix model extension  
 \eqref{limit}. We are lead to define  a characteristic distance
$L_{tail}$ such that for $L=|\xpr|>L_{tail}(s)$ the impact-parameter
scattering amplitude is given by Eq.~\eqref{e.impact}.
One can then divide the whole impact-parameter space into 
a $tail$ region ($L>L_{tail}$), and a $core$ region ($L<L_{tail}$)
where inelastic channels are supposed to open up.
More specifically, the following regions are identified (see Fig.~\ref{2}). 
\begin{enumerate}
\item At  large distances $L > L_{max}$, whose  exact expression is given by
  \eqref{e.constraint}, the gravitational field in the bulk is weak enough,
  and the contribution of the disconnected minimal surface gives a rigorous
  holographic determination of the impact-parameter tail of the
  scattering amplitude (Fig.~\ref{2}-1).
\item At moderately large distances $L_{min}<L<L_{max}$, where $L_{min}$
has been defined in \eqref{limit}, the strong gravitational field is expected
to generate a non zero $\Im\,\dl_G$ leading to inelastic contributions on the
gauge theory side. 
The minimal surface is still disconnected but the gravitational field
begins to become strong in some relevant region in the bulk.
Nevertheless, for the sake of completeness, we will investigate what happens
assuming the validity of the elastic eikonal expression up to $L_{tail}$, lying somewhere 
in the range $L_{min}\le L_{tail}\le L_{max}$ (Fig.~\ref{2}-2).
\item For $L_{connect}<L<L_{min}$ the elastic eikonal expression
  \eqref{e.impact} is no more reliable, even from the $S$-matrix
  point-of-view. An eikonal formula may still be valid with an  imaginary
  contribution to the phase shifts but it cannot be obtained through  the weak
  gravity regime of the AdS/CFT correspondence, even if the minimal  surface
  is still made of  disconnected surfaces joined by interacting 
  fields (Fig.~\ref{2}-3). 
\item Finally, for even smaller distances $L\le L_{connect}$ the Gross-Ooguri
 transition~\cite{Gross} takes place, and the minimal surface solution
 becomes connected. In this region, the AdS/CFT description goes
 beyond the interaction mediated by supergravity fields (Fig.~\ref{2}-4).
\end{enumerate}
Region 1 and possibly part of region 2 constitute the impact-parameter
{\it tail}  
region, while regions 3 and 4 constitute the central impact-parameter
{\it core}  
region. 
Since we know precisely $a(\chi,\xpr)$ only in the $tail$
region, we are able to determine only part of the full scattering 
amplitude, $i.e.$, the large impact-parameter contribution ${\cal A }_{tail}$,
\eq
{\cal A }\equiv {\cal A }_{core}+{\cal A }_{tail}\ ;\quad{\cal A }_{tail}(s,t;\vec{R}_1,\vec{R}_2) = 2is \int_{L\ge L_{tail}}
d^2\xpr\ e^{i{\vec q}\cdot\xpr}\,\left[1-e^{\left(i\sum_\psi \dl_\psi\right)}\right]\ ;
\label{Tail}
\eqx
nevertheless, constraining ${\cal A }_{core}$ with the unitarity bound
for the impact-parameter amplitude $\Im\,a(\chi,\xpr)\le 2$, 
we will be able to set a lower and
an upper bound on the large-$s$ behavior of the total cross section. 

\section{Total cross sections at high energy in $\nnn=4$ SYM}

\begin{figure}[t]
  \centering
  \includegraphics[width=0.5\textwidth]{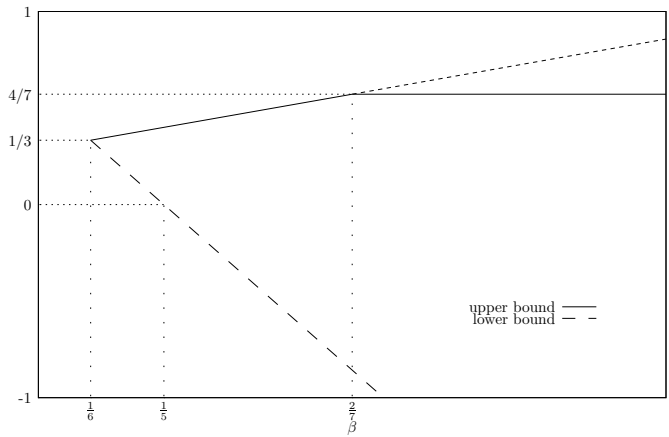}
\caption{{\it Upper and lower bounds on the high-energy behavior of
    total cross  sections.} The bounds on the exponent $\gamma$ of the
  total cross section $\sigma_{tot}\propto \s^\gamma$,
  $i.e.$, on the Pomeron intercept minus one, are displayed as a
  function of the power law exponent $\bt$ of  
  $L_{tail}\propto \s^{\bt}$. {\it Solid line}: upper bound, coming from
  the core contribution for $\f{1}{6}\le\bt\le\f 27$.  {\it Long-dashed line}: lower
  bound, coming from the tail contribution for $\bt\ge\f{1}{6}$. 
{\it Short-dashed line}: weaker upper bound for $\bt > \f 27$,
obtained by overestimating the $core$ contribution (see text).  
}
\label{3}
\end{figure}

In order to evaluate the contribution $\sigma_{tail}$ to the total
cross section of the large impact-parameter region, as obtained from
AdS/CFT, we need only the imaginary part of the amplitude at $t=0$,
which is related to the dipole-dipole total cross section by
means of the optical theorem. We can then ignore the divergence in the
real part, due to the KK scalar exchange, discussed in~\cite{adsbounds}. We have  
\eq
\begin{aligned}
  \sigma_{tail}& {\mathop \simeq_{s\to\infty}}  \f{\Im \, {\cal A}_{tail}(s,0;\vec{R}_1,\vec{R}_2)}{s} =
  4\pi\int_{L_{tail}}^{\infty}\!\!
  dL\,L\, \bigg[1-\cos\bigg(\sum_\psi \dl_\psi\bigg)\bigg].
\label{total}
\end{aligned}
\eqx
The $\chi$-dependence  at large energy induces a
hierarchy  between the different contributions, 
clearly revealed after performing the change of
variables  $\lam \equiv   L/(\sinh\chi)^{\f{1}{6}}{\sqrt{R_1R_2}}$,
$\lam_{tail} \equiv   L_{tail}/(\sinh\chi)^{\f{1}{6}}{\sqrt{R_1R_2}}$,  
which yields
\eq 
\begin{aligned}
  \sigma_{tail} =& 4\pi(\sinh\chi)^{\f{1}{3}} R_1R_2
 \int_{\lam_{tail}}^{\infty} \! d\lam\,\lam 
 \Bigg[1-\cos\Bigg(\f{\ka_S}{\lam^{2}}\frac{1}{(\sinh\chi)^{\f{4}{3}}} +
 \f{\ka_D}{\lam^{6}}\frac{1}{(\sinh\chi)^2} +
   \f{\ka_B}{\lam^{4}}\frac{\coth\chi}{(\sinh\chi)^{\f{2}{3}}} + 
\f{\ka_G}{\lam^{6}}(\coth\chi)^2\Bigg)\Bigg]
\label{integrallam}
\end{aligned}
\eqx 
(rescaling with $\sinh\chi$ instead of
$\cosh\chi$ allows to keep manifest the symmetry under crossing, $i.e.$, under 
$\chi\to i\pi-\chi$, of the various phase shifts). 
Note that the leading term, coming from graviton
exchange, is crossing-symmetric, thus  
corresponding to ``Pomeron exchange'' in the $S$-matrix language, while the
first subleading term, coming from 
antisymmetric-tensor exchange, 
 is crossing-antisymmetric, thus corresponding to
``Odderon exchange''. 
At large energy, retaining only the dominant contribution, we have
\eq
\begin{aligned}
 \sigma_{tail}  &\mathop \simeq_{s\to\infty} 4\pi {R}_{1}{R}_{2}
  \s^{\frac{1}{3}}\int_{\lam_{tail}}^{\infty} d\lambda \,\lambda
  \left[1-\cos\left(\frac{\ka_G}{\lambda^6}\right)\right]
  \,\,\, = \,\,\, \f{2\pi}{3} {R}_{1}{R}_{2}\,\left(\f{\s}{\mu_{tail}}\right)^{\f{1}{3}}
  \int_0^{1} dx\, x^{-\f{4}{3}}
  \left[1-\cos\left(\ka_G\mu_{tail}x\right)\right]\  ,
\label{mu}
\end{aligned}
\eqx
where we have set 
 $\mu_{tail}= \lam^{-{6}}_{tail}$, 
 $x=\f{\lam^{-6}}{\mu_{tail}}$. 
To complete the
calculation of the high-energy behavior of $\sigma_{tail}$ we need
to know the limit of validity of the eikonal expression, and thus 
how $L_{tail}$ depends on $s$. Let us consider the parameterization 
\eq
L_{tail}=\lam_0 \sqrt{R_1R_2}
\ \varsigma^\beta\Rightarrow \quad \lam_{tail}=\lam_0\ 
\varsigma^{\beta-\f{1}{6}}\ ,\quad\mu_{tail}=\lam_0^{-6}
\varsigma^{1-6\beta}\ ,
\label{para}
\eqx
where $\lam_0$ may have some residual dependence on $R_{1,2}$ (see
$e.g.$~\eqref{e.constraint}). 
According to the value of $\beta$ we have for large $s$
\eq
\begin{aligned}
  \label{eq:tail}
  \sigma_{tail} 
&\mathop\sim_{s\to \infty} \f{2\pi}{3}R_1R_2\ \left\{
  \begin{aligned}
    &\s^{\f{1}{3}}\f{3\pi\ka_G^{\f{1}{3}}}{\Gamma(1/3)} & &\beta < \f{1}{6}\ ,\\
    &\s^{\f{1}{3}} \lam_0^2  \int_0^{1} dx\, x^{-\f{4}{3}}
    \left[1-\cos\left(\ka_G\lam_0^{-6}x\right)\right]     & & \beta=\f{1}{6}\ ,\\
    & \s^{2-10\beta}\frac{1}{2}\ka_G^2\lam_0^{-10} & & \beta>\f{1}{6}\ .
  \end{aligned}\right.
\end{aligned}
\eqx
We are now in the position to determine a lower and an upper bound on the
high-energy behavior of the dipole-dipole total cross section. 
Since obviously $\sigma_{tot}>\sigma_{tail}$, Eq.~\eqref{eq:tail} provides a
{\it lower} bound. The overall unitarity constraint 
allows
one to put an {\it upper} bound on the contribution from the $core$ region
$L<L_{tail}$, $i.e.$, $\sigma_{core} \le 4\pi L_{tail}^2 =
4\pi\lam_0^2R_1R_2\s^{2\beta}$, and thus on the whole total cross
section. The bounds can be written in a compact way as 
\begin{equation}
  \label{eq:bound}
  \min\left(\f{1}{3},2-10\beta\right) \le \lim_{\s\to\infty}\f{\log \sigma_{tot}}{\log\s} \le
  \max\left(\f{1}{3},2\beta\right),
\end{equation}
and they are shown in Fig.~\ref{3}. In particular, using the value
$\beta=\f{2}{7}$ coming from the weak field 
constraint \eqref{e.constraint}, one obtains the rigorous bound $-6/7 \le
\lim_{\s\to\infty} {\log \sigma_{tot}}/{\log\s} \le 4/7$. 
The following remarks are in order.
\begin{enumerate}
\item For $\beta<\f{1}{6}$, at sufficiently high energy one would 
have $L_{tail}<L_{min}$, thus entering the unphysical region where the impact-parameter 
partial amplitude is infinitely oscillating; moreover, the total cross 
section would become purely elastic at high energy, while one expects the opening 
of more and more inelastic channels as the energy increases: this means that one 
lies beyond the applicability of the elastic eikonal approximation. 
\item At $\beta=\f{1}{6},$ corresponding to $L_{tail}/L_{min}=const.$,  the $tail$ and $core$ contributions have the same
  high-energy behavior. In this case $\lam_{tail}$ does not depend on energy. However, one has to verify the 
  non-oscillating behavior condition $\lam_{tail}\ge
  \left({\ka_G}/\pi\right)^{\f16}$. In this case the bounds transform
  into a prediction,  $\sigma_{tot}\sim  \s^{\f 13}$ (see also~\cite{LP}).
\item For $\f16<\bt\le\f{2}{7}$, which corresponds to $L_{min}<L_{tail}<L_{max}$ (strictly speaking, at sufficiently high energy), 
  the $core$ region dominates, while the $tail$ region gives a
  subleading contribution as $s\to\infty$. The two bounds determine a
  window of possible power-law behaviors. 
\item For the maximal value  $\beta= \f 27$, $i.e.$, for $L_{tail}/L_{max}=const.$, the total
  cross section behavior is constrained to be such that $\sigma_{tot}\le { const.}\times\s^{\f 47}.$ 
  This maximal value is determined from the requirement that the AdS/CFT correspondence can be
  reliably applied, expressed through the constraint \eqref{e.constraint}. In fact, this is the rigorous result obtained 
  by means of the AdS/CFT correspondence, since for smaller $\bt$ one expects inelastic contributions
  coming from a strong dual gravitational field. Note that it
  restricts the total cross section to be below the $bare$ graviton
  exchange contribution, namely $\sigma_{tot}\sim \s^1$.
\item One could also consider $\bt>\f 27$, but in that case one would only obtain a weaker bound on the total cross section. Indeed, in doing so one
  would overestimate the contribution of the $core$, including in it the impact-parameter region $L_{max}<L<L_{tail}$, where the amplitude is reliably
  described by the eikonal AdS/CFT expression.
\end{enumerate}
The absolute bound we obtain is  linked to the
precise derivation of a weak gravitational field limitation of the
AdS/CFT correspondence in the supergravity formulation. Our result
appears as the analogue of the Froissart bound, but in 
the context of the non confining $\nnn=4$ SYM theory, since it is the
combination of the unitarity bound on impact-parameter amplitudes with
the determination of a precise power-like bound on the
impact-parameter radius from AdS/CFT. 
A more stringent bound would be obtained if one assumed
the validity of the elastic eikonal approximation in a region with
strong gravitational field in the bulk; 
however, in this region other contributions are expected to modify the
gravitational sector.   

\begin{acknowledgements}
  I would like to thank the organizers of the ``Low x meeting'', and
  moreover R.~Peschanski, with whom the results discussed
  here were obtained, and R.~Janik, whose calculations were 
  used in their derivation. I would like to thank also the
  IPhT-CEA/Saclay for hospitality. This work has been partly funded by
  a grant of the ``Fondazione Angelo Della Riccia'' (Firenze, Italy). 
\end{acknowledgements}

\end{document}